\documentstyle[10pt,epsf,epsfig,dp_delphititle,rotating]{dp_delphi}
\makeindex
\def\DpPaperGroup{PH-EP}
\def\DpPaperRef{2006-016}
\def\DpDate{{16 May 2006}}
\def\DpAuthors{DELPHI Collaboration}
\def\DpTitle{{Masses, Lifetimes and Production Rates
              of \boldmath $\Xi^-$ and $\overline{\Xi}^+$ at LEP 1}}
\def\DpSubmit{(Accepted by  Physics Letters B)}
\def\DpComment{ }
\def\DpEMail{ }

\newcommand{\epem} {\ifmmode{e^+e^-}\else{$e^+e^-$}\fi}
\newcommand{\qqbar} {\ifmmode{q\bar{q}}\else{$q\bar{q}$}\fi}

\newcommand{\eps}{{\ifmmode \varepsilon \else $\varepsilon$\fi}}
\newcommand{\thetac}{{\ifmmode \theta_C\else $\theta_C$\fi}}

\newcommand{\BC}{\begin{center}}
\newcommand{\EC}{\end{center}}
\newcommand{\BE}{\begin{equation}}
\newcommand{\EE}{\end{equation}}
\newcommand{\BEA}{\begin{eqnarray}}
\newcommand{\EEA}{\end{eqnarray}}
\newcommand{\BA}{\begin{array}}
\newcommand{\EA}{\end{array}}
\newcommand{\BI}{\begin{itemize}}
\newcommand{\EI}{\end{itemize}}
\newcommand{\BF}{\begin{figure}}
\newcommand{\EF}{\end{figure}}
\newcommand{\BT}{\begin{table}}
\newcommand{\ET}{\end{table}}
\newcommand{\BTB}{\begin{tabular}}
\newcommand{\ETB}{\end{tabular}}
\newcommand\BM{\begin{minipage}}
\newcommand\EM{\end{minipage}}

\newcommand{\Erg}[3]{\ifmmode{#1\pm#2_{stat.}\pm#3_{sys.}}\else{$#1\pm#2_{stat.}\pm#3_{sys.}$}\fi}
\newcommand{\erg}[3]{\ifmmode{\scriptstyle#1\,\pm\,#2\,\pm\,#3}\else{$\scriptstyle#1\,\pm\,#2\,\pm\,#3$}\fi}
\begin{document}
\makeatletter
\newcount\@tempcntc
\def\@citex[#1]#2{\if@filesw\immediate\write\@auxout{\string\citation{#2}}\fi
  \@tempcnta\z@\@tempcntb\m@ne\def\@citea{}\@cite{\@for\@citeb:=#2\do
    {\@ifundefined
       {b@\@citeb}{\@citeo\@tempcntb\m@ne\@citea\def\@citea{,}{\bf ?}\@warning
       {Citation `\@citeb' on page \thepage \space undefined}}%
    {\setbox\z@\hbox{\global\@tempcntc0\csname b@\@citeb\endcsname\relax}%
     \ifnum\@tempcntc=\z@ \@citeo\@tempcntb\m@ne
       \@citea\def\@citea{,}\hbox{\csname b@\@citeb\endcsname}%
     \else
      \advance\@tempcntb\@ne
      \ifnum\@tempcntb=\@tempcntc
      \else\advance\@tempcntb\m@ne\@citeo
      \@tempcnta\@tempcntc\@tempcntb\@tempcntc\fi\fi}}\@citeo}{#1}}
\def\@citeo{\ifnum\@tempcnta>\@tempcntb\else\@citea\def\@citea{,}%
  \ifnum\@tempcnta=\@tempcntb\the\@tempcnta\else
   {\advance\@tempcnta\@ne\ifnum\@tempcnta=\@tempcntb \else \def\@citea{--}\fi
    \advance\@tempcnta\m@ne\the\@tempcnta\@citea\the\@tempcntb}\fi\fi}
 
\makeatother
\begin{titlepage}
\pagenumbering{roman}
\CERNpreprint{\DpPaperGroup}{\DpPaperRef} 
\date{{\small\DpDate}} 
\title{\DpTitle} 
\address{\DpAuthors} 
\begin{shortabs} 
\noindent
%
\newcommand{\wffe}[4]
{\mbox{$#1 \pm #2 \,\, {\rm (stat.)}\,\,\pm #3 \,\, {\rm (syst.)}\,\,{\rm #4}$}}
\newcommand{\sig}{\mbox{$\Sigma^0$}}
\newcommand{\sigb}{\mbox{$\overline{\Sigma}^0$}}
\newcommand{\lam}{\mbox{$\Lambda$}}
\newcommand{\xin}{\mbox{$\Xi^-$}}
\newcommand{\xip}{\mbox{$\overline{\Xi}^+$}}

\noindent

Measurements of the \xin\ and \xip\ masses, mass differences, lifetimes 
and lifetime differences 
are presented. 
The \xip\ sample used is much larger than those used previously for such
measurements.
In addition, the $\Xi$ production rates in Z$\to b\bar b$ 
and Z$\to q\bar q$ events are compared and
the position $\xi^*$ of the maximum of the $\xi$ distribution in 
Z$\to q\bar q$ events is measured.
%
%

\end{shortabs}
\vfill
\begin{center}
\DpSubmit \ \\ 
\DpComment \ \\
\DpEMail \ \\
\end{center}
\vfill
\clearpage
\headsep 10.0pt
\addtolength{\textheight}{10mm}
\addtolength{\footskip}{-5mm}
\begingroup
%
\newcommand{\DpName}[2]{\hbox{#1$^{\ref{#2}}$},\hfill}
\newcommand{\DpNameTwo}[3]{\hbox{#1$^{\ref{#2},\ref{#3}}$},\hfill}
\newcommand{\DpNameThree}[4]{\hbox{#1$^{\ref{#2},\ref{#3},\ref{#4}}$},\hfill}
\newskip\Bigfill \Bigfill = 0pt plus 1000fill
\newcommand{\DpNameLast}[2]{\hbox{#1$^{\ref{#2}}$}\hspace{\Bigfill}}
%
\footnotesize
\noindent
\DpName{J.Abdallah}{LPNHE}
\DpName{P.Abreu}{LIP}
\DpName{W.Adam}{VIENNA}
\DpName{P.Adzic}{DEMOKRITOS}
\DpName{T.Albrecht}{KARLSRUHE}
\DpName{T.Alderweireld}{AIM}
\DpName{R.Alemany-Fernandez}{CERN}
\DpName{T.Allmendinger}{KARLSRUHE}
\DpName{P.P.Allport}{LIVERPOOL}
\DpName{U.Amaldi}{MILANO2}
% ...
\normalsize
\endgroup
\newpage
\titlefoot{Department of Physics and Astronomy, Iowa State
     University, Ames IA 50011-3160, USA
    \label{AMES}}
\titlefoot{Physics Department, Universiteit Antwerpen,
     Universiteitsplein 1, B-2610 Antwerpen, Belgium \\
     \indent~~and IIHE, ULB-VUB,
     Pleinlaan 2, B-1050 Brussels, Belgium \\
     \indent~~and Facult\'e des Sciences,
     Univ. de l'Etat Mons, Av. Maistriau 19, B-7000 Mons, Belgium
    \label{AIM}}
\titlefoot{Physics Laboratory, University of Athens, Solonos Str.
     104, GR-10680 Athens, Greece
    \label{ATHENS}}
\titlefoot{Department of Physics, University of Bergen,
     All\'egaten 55, NO-5007 Bergen, Norway
    \label{BERGEN}}
\titlefoot{Dipartimento di Fisica, Universit\`a di Bologna and INFN,
     Via Irnerio 46, IT-40126 Bologna, Italy
    \label{BOLOGNA}}
\titlefoot{Centro Brasileiro de Pesquisas F\'{\i}sicas, rua Xavier Sigaud 150,
     BR-22290 Rio de Janeiro, Brazil \\
     \indent~~and Depto. de F\'{\i}sica, Pont. Univ. Cat\'olica,
     C.P. 38071 BR-22453 Rio de Janeiro, Brazil \\
     \indent~~and Inst. de F\'{\i}sica, Univ. Estadual do Rio de Janeiro,
     rua S\~{a}o Francisco Xavier 524, Rio de Janeiro, Brazil
    \label{BRASIL}}
\titlefoot{Coll\`ege de France, Lab. de Physique Corpusculaire, IN2P3-CNRS,
     FR-75231 Paris Cedex 05, France
    \label{CDF}}
\titlefoot{CERN, CH-1211 Geneva 23, Switzerland
    \label{CERN}}
\titlefoot{Institut de Recherches Subatomiques, IN2P3 - CNRS/ULP - BP20,
     FR-67037 Strasbourg Cedex, France
    \label{CRN}}
\titlefoot{Now at DESY-Zeuthen, Platanenallee 6, D-15735 Zeuthen, Germany
    \label{DESY}}
\titlefoot{Institute of Nuclear Physics, N.C.S.R. Demokritos,
     P.O. Box 60228, GR-15310 Athens, Greece
    \label{DEMOKRITOS}}
\titlefoot{FZU, Inst. of Phys. of the C.A.S. High Energy Physics Division,
     Na Slovance 2, CZ-180 40, Praha 8, Czech Republic
    \label{FZU}}
\titlefoot{Dipartimento di Fisica, Universit\`a di Genova and INFN,
     Via Dodecaneso 33, IT-16146 Genova, Italy
    \label{GENOVA}}
\titlefoot{Institut des Sciences Nucl\'eaires, IN2P3-CNRS, Universit\'e
     de Grenoble 1, FR-38026 Grenoble Cedex, France
    \label{GRENOBLE}}
\titlefoot{Helsinki Institute of Physics and Department of Physical Sciences,
     P.O. Box 64, FIN-00014 University of Helsinki, 
     \indent~~Finland
    \label{HELSINKI}}
\titlefoot{Joint Institute for Nuclear Research, Dubna, Head Post
     Office, P.O. Box 79, RU-101 000 Moscow, Russian Federation
    \label{JINR}}
\titlefoot{Institut f\"ur Experimentelle Kernphysik,
     Universit\"at Karlsruhe, Postfach 6980, DE-76128 Karlsruhe,
     Germany
    \label{KARLSRUHE}}
\titlefoot{Institute of Nuclear Physics PAN,Ul. Radzikowskiego 152,
     PL-31142 Krakow, Poland
    \label{KRAKOW1}}
\titlefoot{Faculty of Physics and Nuclear Techniques, University of Mining
     and Metallurgy, PL-30055 Krakow, Poland
    \label{KRAKOW2}}
\titlefoot{Universit\'e de Paris-Sud, Lab. de l'Acc\'el\'erateur
     Lin\'eaire, IN2P3-CNRS, B\^{a}t. 200, FR-91405 Orsay Cedex, France
    \label{LAL}}
\titlefoot{School of Physics and Chemistry, University of Lancaster,
     Lancaster LA1 4YB, UK
    \label{LANCASTER}}
\titlefoot{LIP, IST, FCUL - Av. Elias Garcia, 14-$1^{o}$,
     PT-1000 Lisboa Codex, Portugal
    \label{LIP}}
\titlefoot{Department of Physics, University of Liverpool, P.O.
     Box 147, Liverpool L69 3BX, UK
    \label{LIVERPOOL}}
\titlefoot{Dept. of Physics and Astronomy, Kelvin Building,
     University of Glasgow, Glasgow G12 8QQ
    \label{GLASGOW}}
\titlefoot{LPNHE, IN2P3-CNRS, Univ.~Paris VI et VII, Tour 33 (RdC),
     4 place Jussieu, FR-75252 Paris Cedex 05, France
    \label{LPNHE}}
\titlefoot{Department of Physics, University of Lund,
     S\"olvegatan 14, SE-223 63 Lund, Sweden
    \label{LUND}}
\titlefoot{Universit\'e Claude Bernard de Lyon, IPNL, IN2P3-CNRS,
     FR-69622 Villeurbanne Cedex, France
    \label{LYON}}
\titlefoot{Dipartimento di Fisica, Universit\`a di Milano and INFN-MILANO,
     Via Celoria 16, IT-20133 Milan, Italy
    \label{MILANO}}
\titlefoot{Dipartimento di Fisica, Univ. di Milano-Bicocca and
     INFN-MILANO, Piazza della Scienza 2, IT-20126 Milan, Italy
    \label{MILANO2}}
\titlefoot{IPNP of MFF, Charles Univ., Areal MFF,
     V Holesovickach 2, CZ-180 00, Praha 8, Czech Republic
    \label{NC}}
\titlefoot{NIKHEF, Postbus 41882, NL-1009 DB
     Amsterdam, The Netherlands
    \label{NIKHEF}}
\titlefoot{National Technical University, Physics Department,
     Zografou Campus, GR-15773 Athens, Greece
    \label{NTU-ATHENS}}
\titlefoot{Physics Department, University of Oslo, Blindern,
     NO-0316 Oslo, Norway
    \label{OSLO}}
\titlefoot{Dpto. Fisica, Univ. Oviedo, Avda. Calvo Sotelo
     s/n, ES-33007 Oviedo, Spain
    \label{OVIEDO}}
\titlefoot{Department of Physics, University of Oxford,
     Keble Road, Oxford OX1 3RH, UK
    \label{OXFORD}}
\titlefoot{Dipartimento di Fisica, Universit\`a di Padova and
     INFN, Via Marzolo 8, IT-35131 Padua, Italy
    \label{PADOVA}}
\titlefoot{Rutherford Appleton Laboratory, Chilton, Didcot
     OX11 OQX, UK
    \label{RAL}}
\titlefoot{Dipartimento di Fisica, Universit\`a di Roma II and
     INFN, Tor Vergata, IT-00173 Rome, Italy
    \label{ROMA2}}
\titlefoot{Dipartimento di Fisica, Universit\`a di Roma III and
     INFN, Via della Vasca Navale 84, IT-00146 Rome, Italy
    \label{ROMA3}}
\titlefoot{DAPNIA/Service de Physique des Particules,
     CEA-Saclay, FR-91191 Gif-sur-Yvette Cedex, France
    \label{SACLAY}}
\titlefoot{Instituto de Fisica de Cantabria (CSIC-UC), Avda.
     los Castros s/n, ES-39006 Santander, Spain
    \label{SANTANDER}}
\titlefoot{Inst. for High Energy Physics, Serpukov
     P.O. Box 35, Protvino, (Moscow Region), Russian Federation
    \label{SERPUKHOV}}
\titlefoot{J. Stefan Institute, Jamova 39, SI-1000 Ljubljana, Slovenia
     and Laboratory for Astroparticle Physics,\\
     \indent~~University of Nova Gorica, Kostanjeviska 16a, SI-5000 Nova Gorica, Slovenia, \\
     \indent~~and Department of Physics, University of Ljubljana,
     SI-1000 Ljubljana, Slovenia
    \label{SLOVENIJA}}
\titlefoot{Fysikum, Stockholm University,
     Box 6730, SE-113 85 Stockholm, Sweden
    \label{STOCKHOLM}}
\titlefoot{Dipartimento di Fisica Sperimentale, Universit\`a di
     Torino and INFN, Via P. Giuria 1, IT-10125 Turin, Italy
    \label{TORINO}}
\titlefoot{INFN,Sezione di Torino and Dipartimento di Fisica Teorica,
     Universit\`a di Torino, Via Giuria 1,
     IT-10125 Turin, Italy
    \label{TORINOTH}}
\titlefoot{Dipartimento di Fisica, Universit\`a di Trieste and
     INFN, Via A. Valerio 2, IT-34127 Trieste, Italy \\
     \indent~~and Istituto di Fisica, Universit\`a di Udine,
     IT-33100 Udine, Italy
    \label{TU}}
\titlefoot{Univ. Federal do Rio de Janeiro, C.P. 68528
     Cidade Univ., Ilha do Fund\~ao
     BR-21945-970 Rio de Janeiro, Brazil
    \label{UFRJ}}
\titlefoot{Department of Radiation Sciences, University of
     Uppsala, P.O. Box 535, SE-751 21 Uppsala, Sweden
    \label{UPPSALA}}
\titlefoot{IFIC, Valencia-CSIC, and D.F.A.M.N., U. de Valencia,
     Avda. Dr. Moliner 50, ES-46100 Burjassot (Valencia), Spain
    \label{VALENCIA}}
\titlefoot{Institut f\"ur Hochenergiephysik, \"Osterr. Akad.
     d. Wissensch., Nikolsdorfergasse 18, AT-1050 Vienna, Austria
    \label{VIENNA}}
\titlefoot{Inst. Nuclear Studies and University of Warsaw, Ul.
     Hoza 69, PL-00681 Warsaw, Poland
    \label{WARSZAWA}}
\titlefoot{Fachbereich Physik, University of Wuppertal, Postfach
     100 127, DE-42097 Wuppertal, Germany \\
\noindent
{$^\dagger$~deceased}
    \label{WUPPERTAL}}
\addtolength{\textheight}{-10mm}
\addtolength{\footskip}{5mm}
\clearpage
\headsep 30.0pt
\end{titlepage}
%
\pagenumbering{arabic} 
\setcounter{footnote}{0} %
\large
%
%


\newcommand{\xin}{\mbox{$\Xi^-$}}
\newcommand{\xip}{\mbox{$\overline{\Xi}^+$}}
\newcommand{\zz}{\mbox{$Z$}}
\newcommand{\wffe}[3]
{\mbox{$#1 \pm #2 \,\, {\rm (stat.)}\,\,\pm #3 \,\, {\rm (syst.)}$}}
\newcommand{\qq}{\mbox{$q \bar q$}}
\newcommand{\sig}{\mbox{$\Sigma^0$}}
\newcommand{\sigb}{\mbox{$\overline{\Sigma}^0$}}
\newcommand{\lam}{\mbox{$\Lambda$}}
\newcommand{\pio}{\mbox{$\pi^0$}}
\newcommand{\pythia}[1]{{\sc pythia} #1}
\newcommand{\jetset}[1]{{\sc jetset} #1}
\newcommand{\herwig}[1]{{\sc herwig} #1}
\newcommand{\delsim}{{\sc delsim}}
\newcommand{\wff}[4]{#1 $\pm$ #2 (stat.) $\pm$ #3 (syst.) #4}
\newcommand{\wffx}[3]{#1 $\pm$ #2 (stat.) $\pm$ #3 (syst.)}

\renewcommand{\thesection}{\arabic{section}}
\renewcommand{\thesubsection}{\thesection.\arabic{subsection}}
\renewcommand{\thefootnote}{\fnsymbol{footnote}}    
\setcounter{footnote}{1}                            %
\renewcommand{\thesection}{\arabic{section}}
\renewcommand{\arraystretch}{1.2}
\section{Introduction}

This paper presents measurements of the masses and mean
lifetimes of {\xin} and {\xip} and of their mass and lifetime differences,
together with a study of \xin 
\footnote{Antiparticles are implicitly included unless explicitly stated
otherwise.} 
production in $Z^0$ hadronic decays.

Previous measurements of 
the {\xip} mass and mean lifetime suffer from low statistics compared to {\xin}
measurements, since they came from 
bubble chamber or hyperon beam experiments 
with a large asymmetry in the production of {\xin} and \xip.
The Particle Data Group~\cite{PDG} lists only $\sim$ 80 events 
used for measurement of the {\xip} mass
and 34 for its mean lifetime, compared to $\sim$ 2400 events for the 
\xin\ mass and $\sim$ 87000 for its mean lifetime. 
The present analysis  uses about 2500 {\xin} and 2300 {\xip},
with small backgrounds.
The symmetry in the production of particles and 
antiparticles in $Z^0$ decays makes direct measurements
of {\xin} and {\xip} mass and lifetime differences with high precision
feasible. 
A non-zero value of either difference would signal violation of CPT invariance.


A comparison of the $\Xi$ production rates in
$Z^0 \rightarrow b\bar b$ and $Z^0 \rightarrow q\bar q$ events 
is also presented, 
together with a measurement of the position $\xi^*$ 
of the maximum of the distribution in $\xi = -\ln x_p$, 
where $x_p$ is the fractional $\Xi$ momentum.

\section{The DELPHI detector and event selection}

The DELPHI detector is described elsewhere~\cite{NIM1,NIM2}.
The detectors most important for this analysis 
are the Vertex Detector (VD), the Inner Detector (ID),
the Time Projection Chamber (TPC), and the Outer Detector (OD).
The VD consists of three concentric layers of silicon strip detectors,
located at radii of 6~cm, 9~cm and 11~cm.
The data used here were taken in 1992-1995 inclusive, when
the polar angles covered for particles crossing all three VD layers were
$43 ^\circ < \theta < 137 ^\circ$, where $\theta$ is given with respect to
the $z$ axis\footnote{In 
the standard DELPHI coordinate system, the $z$ axis is along the electron
direction, the $x$ axis points towards the centre of LEP, and the $y$ axis
points upwards. The polar angle to the $z$ axis is called $\theta$ and the
azimuthal angle around the $z$ axis is called $\phi$; the radial coordinate is
$R = \sqrt{x^2+y^2}$.}.
In 1994 and 1995, the first and third layers
had double-sided readout and gave both $R\phi$ and $z$ coordinates.
The TPC is the main tracking device where charged-particle tracks are 
reconstructed in three dimensions for radii between 29~cm and 122~cm.
The ID and OD are two drift chambers located at radii between 12~cm and 28~cm
and between 198~cm and 206~cm respectively, and
provide additional points for the track reconstruction.


A charged particle was accepted in the analysis if its
track length was above 30 cm,
its momentum above 100 MeV/c, and
its relative momentum error below 100\%.


An event was classified as hadronic if it had at least 7 charged particles
with momentum above 200 MeV/c carrying 
more than 15 GeV reconstructed energy in total 
and at least 3 GeV in each hemisphere defined with respect to the $z$ axis.


The analysis used 3.25 million reconstructed hadronic decays of the \zz, consisting of
0.67 million from the 1992 run, 
0.68 million from 1993,
1.29 million from 1994, 
and 0.61 million from 1995.
 

Simulated events were produced using the \jetset parton shower 
generator~\cite{JETSET},
and then processed with the DELPHI event simulation program
\delsim~\cite{NIM2} which fully simulates all detector effects.
For each of the years 1992 to 1994, about 1 million fully simulated
\qq\ events were analyzed in the same way as the real data,
and about 0.6 million for 1995.
The total number of simulated
events used was thus about 3.6 million, comparable to the number of
real events.
The number of {\xin} and {\xip} decays
generated in the simulation was about 89000.

\section{Analysis}

The $\Xi^-$ hyperon was
studied by a complete reconstruction of the decay chain
\mbox{$\Xi^- \rightarrow \Lambda \pi^-$}, where
\mbox{$\Lambda \rightarrow p\pi^-$}. A similar analysis procedure was 
used previously for $\Omega^-$ reconstruction~\cite{omsig}.


All pairs of oppositely-charged particles were tried in a search for $\Lambda$
candidates. For each such pair, a vertex fit was performed by the standard
DELPHI $V^0$ search algorithm\footnote{
A $V^0$ consists of two oppositely charged particles originating from a neutral
particle decaying in flight.
}~\cite{NIM2}. 
A pair was
accepted as a $\Lambda$ candidate if the $\chi^2$-probability of the secondary
vertex fit exceeded 0.1\%, the measured flight distance from the primary vertex
of the $\Lambda$ candidate in the $xy$ plane exceeded twice its
error, and the angle between the momentum vector sum of the two tracks and the
vector joining the primary and secondary vertices was less than 0.1 radians
(the loss of signal due to this cut has been shown to be negligible).  
The inclusive \lam\ reconstruction efficiency was around 19\%~\cite{NIM2}, 
including the 63.9\% branching ratio of 
\mbox{$\Lambda\rightarrow\mathrm{p}\pi^-$}~\cite{PDG}.
The invariant mass of the $\Lambda$ candidate was required to be between
1.105~GeV/c$^2$ and 1.125~GeV/c$^2$.


One by one, the remaining tracks of charged particles
that crossed the $\Lambda$ trajectory in the 
$xy$ plane were then combined with
the $\Lambda$ candidate to form a $\Xi^-$ candidate.
All {\xin} were assumed to
originate from the beam interaction point, which was calculated event by event.


A constrained fit was performed if: 
\begin{itemize}
\item
   the intersection between the $\Lambda$ and the charged particle trajectory
   was more than 8 mm away from the main vertex in the $xy$ plane;
\item
   the $\Lambda$ and charged particle trajectories were less 
   than 7 mm apart in
   the $z$ direction at the point of crossing in the $xy$ plane;
\item
   and the charged particle had an impact parameter with
   respect to the main vertex in the $xy$ plane of at least 0.5 mm.
\end{itemize}

 The fit used was a general least squares fit with kinematical and 
 geometrical constraints applied to each \xin\ candidate. 
 The 16 measured variables in the fit were 
 the five parameters of the helix parameterization of each of the three charged
 particle tracks and the $z$ coordinate of the beam interaction point
 (the $x$ and $y$ coordinates  were so precisely measured that
 they could be taken as fixed). The two unmeasured variables were the decay radii 
 of the $\Xi^-$ and $\Lambda$. The $\Xi^-$ decay point was then determined from
 this $\Xi^-$ decay radius and the $\pi^-$ trajectory while the $\Lambda$ decay
 point was determined from the point on the proton trajectory at the $\Lambda$
 decay radius.  The curved \xin\ track was not measured, but calculated in the fit.  

 Four constraints required the momenta of the $\Xi^-$ and $\Lambda$ at their decay
 points to be
 in the same direction as the trajectory joining their production and decay
 positions, two required the other $\pi^-$ to meet the proton at the $\Lambda$
 decay radius, and the  last (seventh) constrained the $\Lambda$ mass to its
 nominal value (1115.684$\pm$0.006) MeV/c$^2$. 
 For further details concerning the fitting procedure, see \cite{TULIP}.
%
%

The pull distributions of the 16 fitted quantities were all 
approximately normally distributed, with mean 0 within $\pm 0.1$
and standard deviation 1 within $\pm 0.1$,
both for data and for the simulated events.


The following cuts were used to select the {\xin} and {\xip} samples:

\begin{itemize}
\item  the $\chi^2$-probability of the fit had to exceed 1\%;
\item  the $\Xi$ momentum, $p_{\Xi}$, had to fulfill $1.2 < \xi < 4.2$ where 
       $\xi= -\ln x_p$ and $x_p = p_{\Xi}/p_{beam}$;
       this corresponds to $0.015<x_p<0.3$
       or $0.7 <p_{\Xi}< 14$ GeV/c;
\item  the $\Xi$ momentum had to point into the barrel region 
       of the detector ($\mid \cos\theta \mid < 0.85$);
\item  the decay radius of the $\Xi$ in the $xy$ plane had to exceed 2 cm;
\item  the decay radius of the $\Xi$ in the $xy$ plane had to be less 
       than the \lam\ decay radius.
\end{itemize}

\begin{figure}[ht]
\begin{center}
\mbox{\epsfig{figure=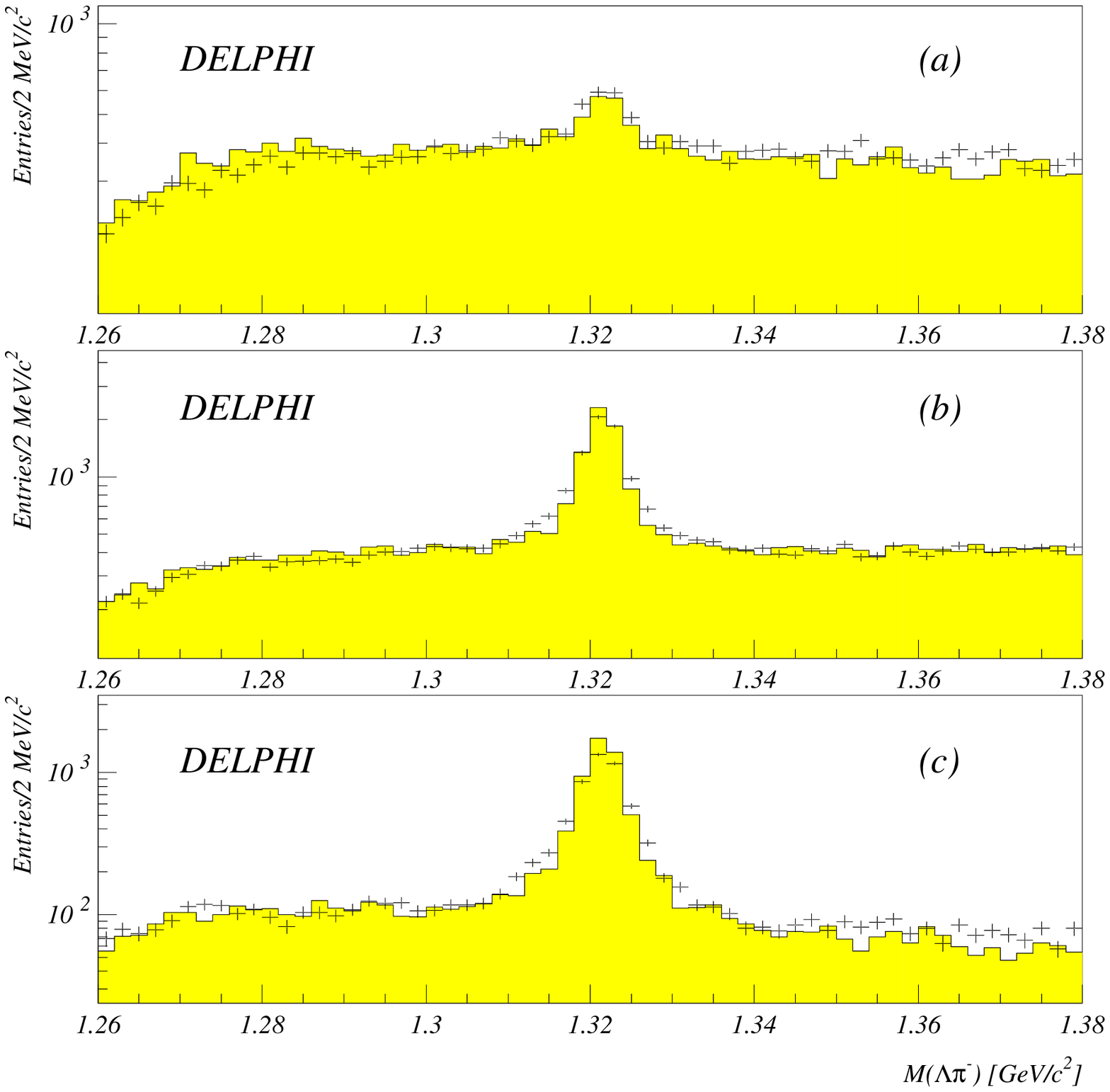,width=\textwidth}}
\end{center}
\caption[blabla]{ 
          The right-sign ($\lam\pi^-$ and $ \overline{\lam}\pi^+$) mass
          distribution with the
          $\Xi$ (\xin\ and \xip\ added) signals in different
          $\chi^2$ probability bins for data and simulation. The data are
          represented by the points with error bars and the simulation by
          the histograms, which are normalized to the same number of entries:
          {\bf (a)} shows the $\Xi$ signal without any other cuts applied 
          for events with a $\chi^2$ fit probability below 1\%,
          {\bf (b)} shows the $\Xi$ signal without any other cuts applied 
          for events with a $\chi^2$ fit probability above 1\%,
          {\bf (c)} shows the $\Xi$ signal after all cuts given in the text 
          were applied
          for events with a $\chi^2$ fit probability above 1\%.}
\label{fig:chi2_bins}
\end{figure}

\begin{figure}[ht]
\begin{center}
\mbox{\epsfig{figure=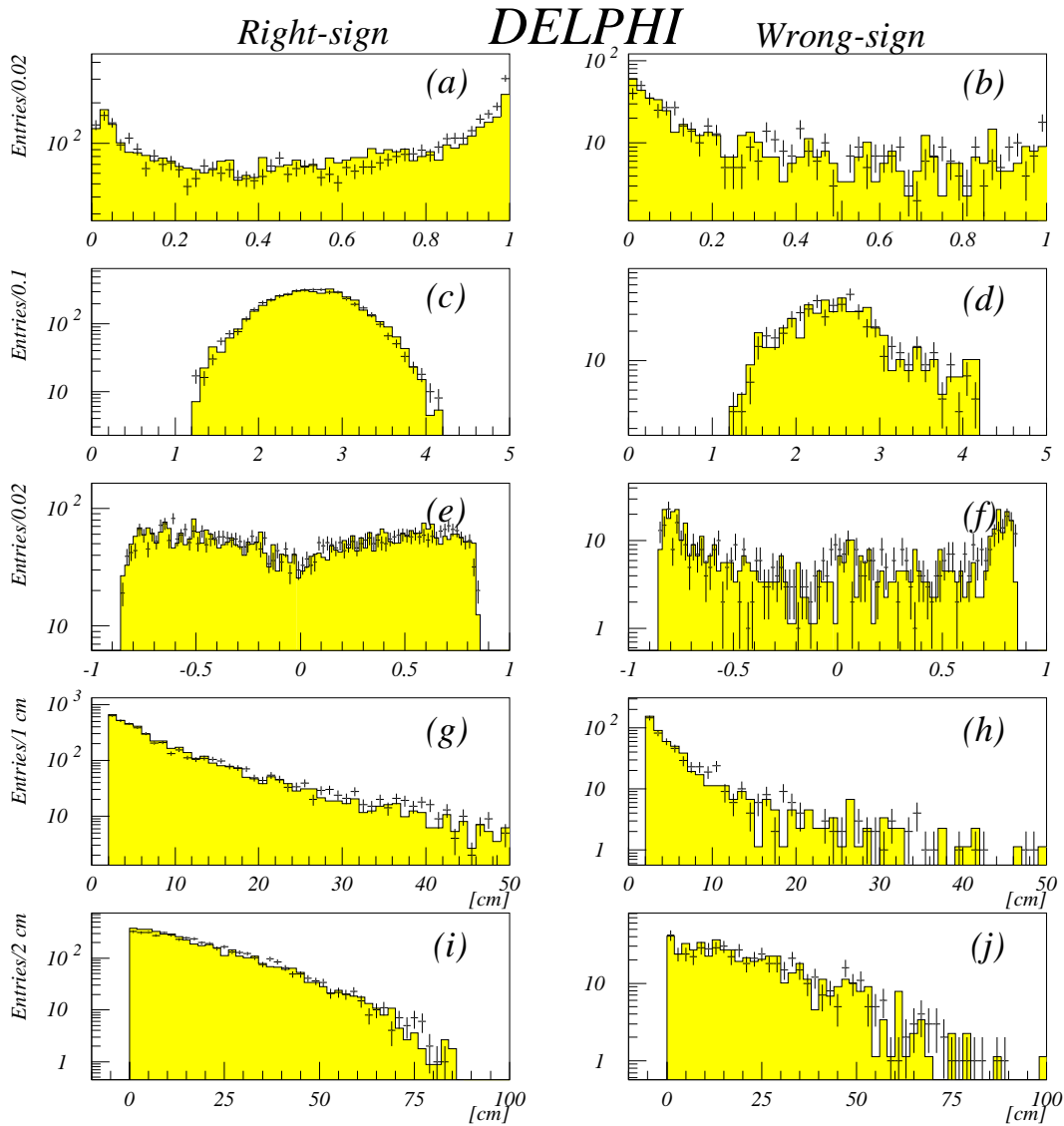,width=\textwidth}}
\end{center}
\caption[blabla]{ 
          All the variables used in the $\Xi$ selection for candidates
          in the mass interval M$_\Xi \pm 5$~MeV/c$^2$. The
          histograms are from the simulation and the points with error bars
          are the data. The years 1992 to 1995 have all been added,
          both for data and simulation. The simulation histograms are
          normalized to the data ones. All variables are shown 
          for right-sign ($\lam\pi^-$ and $ \overline{\lam}\pi^+$)
          and wrong-sign ($\lam\pi^+$ and $ \overline{\lam}\pi^-$) combinations
          after all cuts have been made:
          {\bf (a,b)} $\chi^2$ probability, 
          {\bf (c,d)} $\xi=-\ln(p_{\Xi}/p_{beam})$,
          {\bf (e,f)} cosine of the polar angle $\theta$ of the $\Xi$ momentum,                    
          {\bf (g,h)} flight distance of the $\Xi$ in the $xy$ plane,
          {\bf (i,j)} distance in the $xy$ plane between the $\Lambda$ and
                    $\Xi$ decay points. }
\label{fig:mc_data_xi_1}
\end{figure}

Figure~\ref{fig:chi2_bins} shows the right-sign ($\lam\pi^-$ and
$\overline{\lam}\pi^+$) mass distributions and the $\Xi$ signals before and 
after the cuts were applied.
Apart from a difference in mass resolution, the agreement between data and
simulation was very good.
The distributions
of the variables used in the selection of $\Xi$ candidates 
are shown in Figure~\ref{fig:mc_data_xi_1} for wrong-sign
($\lam\pi^+$ and $ \overline{\lam}\pi^-$) as well as for right-sign
($\lam\pi^-$ and $ \overline{\lam}\pi^+$) combinations.

\begin{figure}
\begin{center}
\mbox{\epsfig{figure=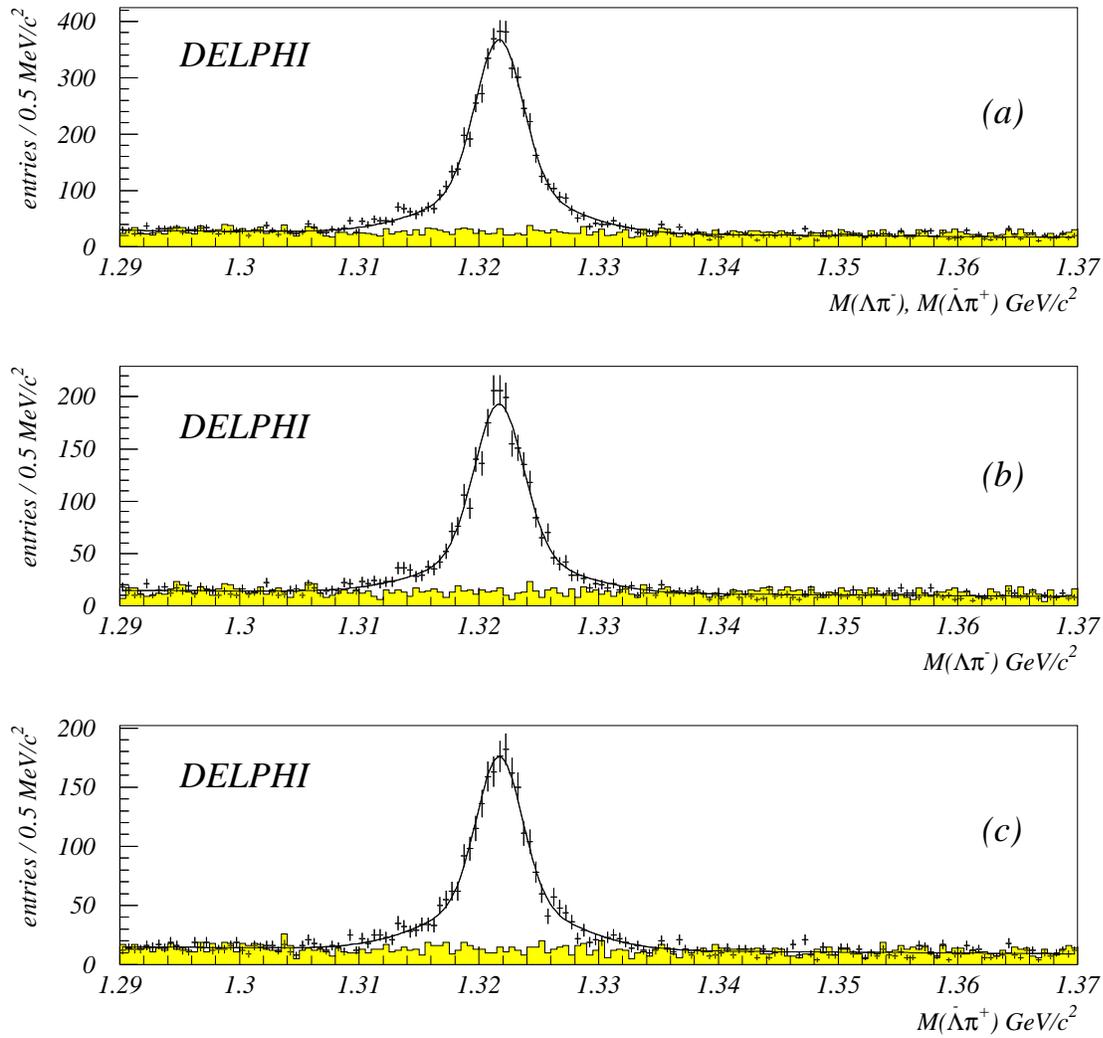,width=\textwidth}} 
\end{center}
\caption{1992-1995 data: {\bf(a)} the {\xin} and {\xip} sample,
{\bf(b)} the {\xin} sample, {\bf(c)} the {\xip} sample,
The points with error bars show the right-sign ($\Lambda\pi^-$,
$\overline{\Lambda}\pi^+$) combinations. The wrong-sign 
($\Lambda\pi^+$, $\overline{\Lambda}\pi^-$)
combinations are shown as the shaded histograms.
The curves show the fits to the $\Xi$ mass distributions 
described in the text
(solid line).}
\label{fig:xi}
\end{figure}

The fit gave a narrow mass peak from $\Xi$ decays on a small background, 
as shown in Figure~\ref{fig:xi}a;
$2478 \pm 68$ {\xin} and $2256 \pm 63$ {\xip} decays 
were reconstructed, as shown in
Figures~\ref{fig:xi}b and \ref{fig:xi}c.
The fitted curves consist of a linear term for the background, and two Gaussian
distributions with common mean for the signal.
The $\Xi$ mass resolution depends on momentum. Therefore the signal is, in
principle, the sum of an infinite number of Gaussians. 
But two give a reasonably good fit.
The fitted widths of the two Gaussians were ($2.0\pm0.1$) MeV/c$^2$ and
($5.6\pm0.4$) MeV/c$^2$,
with a relative fraction of $1.29\pm0.18$. 
The corresponding widths from fitting simulated data were
($1.8\pm0.1$) MeV/c$^2$ and ($5.5\pm0.5$) MeV/c$^2$, with a relative fraction
of $2.01\pm0.27$.
This parameterization of signal and
background was used in the determination of the {\xin} and {\xip} masses. 

The only possible physical background is the decay
\mbox{$\Omega^{\pm} \rightarrow \Lambda {\rm K}^{\pm}$}.
The number of $\Omega^-$ reconstructed in our \xin analysis is estimated to be
at most five, and consequently to have no significant influence. 

\subsection{Measurement of \xin\ and \xip\ masses and mass difference}

Table~\ref{table:ximasses} gives the 
fitted mass and mass difference values for the real data.
As already described,
the signal (see Figure~\ref{fig:xi}) was represented by two Gaussian distributions
with common mean and the background by a linear term.

\begin{table}
\begin{center}
{
\begin{tabular}{|lrrrr|}
\hline
Year  & \multicolumn{1}{c}{92} & \multicolumn{1}{c }{93} &
        \multicolumn{1}{c}{94} & \multicolumn{1}{c|}{95} \\ 
\hline
\hline
M$_{\Xi^-}$ in data
& $1321.60\pm0.17$ & $1321.25\pm0.16$ & $1321.45\pm0.10$ & $1321.50\pm0.16$ \\
M$_{\Xi^+}$ in data
& $1321.70\pm0.18$ & $1321.49\pm0.14$ & $1321.46\pm0.12$ & $1321.19\pm0.18$ \\
M$_{\Xi^{\pm}}$ in data
& $1321.65\pm0.13$ & $1321.37\pm0.11$ & $1321.45\pm0.08$ & $1321.36\pm0.12$ \\
\hline
M$_{\Xi^-}-$M$_{\Xi^+}$ in data 
& $-0.10\pm0.25$ & $-0.23\pm0.22$ & $-0.02\pm0.15$ & $0.31\pm0.24$ \\
\hline
\hline
M$_{\Xi^{\pm}}-$1321.3 in MC 
& $0.14\pm0.07$ & $-0.02\pm0.07$ & $0.06\pm0.07$ & $0.30\pm0.09$ \\
\hline
\hline
Corrected M$_{\Xi^-}$:
& $1321.46\pm0.18$ & $1321.27\pm0.17$ & $1321.39\pm0.12$ & $1321.20\pm0.19$ \\
Corrected M$_{\Xi^+}$:
& $1321.56\pm0.19$ & $1321.51\pm0.16$ & $1321.40\pm0.14$ & $1320.89\pm0.20$ \\
Corrected M$_{\Xi^{\pm}}$:
& $1321.51\pm0.15$ & $1321.39\pm0.12$ & $1321.39\pm0.10$ & $1321.06\pm0.15$ \\
\hline
\end{tabular}
}
\end{center}
\caption{\label{table:ximasses} \xin\ and \xip\ mass fit results. 
Values are in MeV/c$^2$. In the simulated (`MC') sample, the
generated {\xin} mass was 1321.3 MeV/c; the corresponding mass
shifts per year are used to correct the mass values found in the data.
The errors are statistical only.}
\end{table}


In order to correct for any bias due to the data processing 
or to the analysis and fit procedure,
the mass values obtained from the data were corrected by the difference 
between the values obtained in the same way from the simulated events 
and the input value used in the simulation (1321.3 MeV/c$^2$). 
As no effect could be identified that might affect the \xin and \xip masses
differently,
the correction was calculated once for each year, using the
corresponding fully simulated $\Xi^\pm$ sample.
Table~\ref{table:ximasses} also shows these corrections, 
and the corrected mass values. 
The statistical errors of the corrected values contain the statistical
errors of the simulation.

The $\Xi^\pm$ mass value averaged over all years was
($1321.45\pm0.05$) MeV/c$^2$
with a $\chi^2$ probability for the combination of 33\% before correction,
and ($1321.35\pm0.06$) MeV/c$^2$ with a $\chi^2$ probability of 17\%  after
correction.
Thus the average correction amounted to ($-0.10\pm0.04$) MeV/c$^2$.

\subsubsection{Mass scale calibration}
The mass scale 
was calibrated by determining the \lam\ and K$^0_s$ masses in the same way,
and comparing the resulting values with the known values~\cite{PDG}.
The \lam\ and K$^0_s$ samples used for this purpose 
were spread over each whole year and 
their sizes were restricted to 
make it possible to use the same signal and background parameterizations as for
the $\Xi$\footnote{
The reconstructed \lam\ samples were typically twice as large as the
reconstructed $\Xi$ samples while the K$^0_s$ samples were typically 10 times
larger}.

The \lam\ and K$^0_s$ decays were reconstructed by considering all pairs of
oppositely charged particles, and the vertex defined by each pair was determined
by minimizing the $\chi^2$ of the extrapolated tracks.
Consequently, this was a purely geometrical vertex fit, as opposed to the
mass-constrained fit described above for the $\Xi$ candidates.
The measured \lam\ and K$^0_s$ mass offsets from their nominal values are shown
in Table~\ref{table:v0masses}.

\begin{table}
\begin{center}
{
\begin{tabular}{|lrrrr|}
\hline
Year  & \multicolumn{1}{c}{92} & \multicolumn{1}{c }{93} &
        \multicolumn{1}{c}{94} & \multicolumn{1}{c|}{95} \\ 
\hline
\hline
M$_{K^0}$ offset in data
& $-0.87\pm0.06$ & $-1.09\pm0.05$ & $-0.75\pm0.06$ & $-0.80\pm0.06$ \\
M$_{K^0}$ offset in MC
& $0.01\pm0.05$ & $0.56\pm0.04$ & $0.38\pm0.04$ & $0.68\pm0.04$ \\
\hline
Corrected M$_{K^0}$ offset
& $-0.88\pm0.09$ & $-1.65\pm0.08$ & $-1.13\pm0.09$ & $-1.48\pm0.09$ \\
\hline
\hline
M$_{\Lambda}$ offset in data
& $0.14\pm0.05$ & $-0.14\pm0.05$ & $-0.09\pm0.06$ & $-0.07\pm0.06$ \\
M$_{\Lambda}$ offset in MC
& $0.09\pm0.06$ & $0.01\pm0.04$ & $0.09\pm0.04$ & $0.17\pm0.04$ \\
\hline
Corrected M$_{\Lambda}$ offset
& $0.04\pm0.08$ & $-0.15\pm0.07$ & $-0.18\pm0.08$ & $-0.24\pm0.08$ \\
\hline
\hline
Calculated M$_{\Xi}$ offset
& $-0.13\pm0.09$ & $-0.44\pm0.07$ & $-0.36\pm0.06$ & $-0.47\pm0.07$ \\
\hline
\hline
Resulting M$_{\Xi^\pm}$ & $1321.64\pm0.17$ & $1321.82\pm0.14$ &
$1321.75\pm0.12$ & $1321.53\pm0.17$ \\
\hline
\end{tabular}
}
\end{center}
\caption{\label{table:v0masses} 
Measured offsets from the nominal K$^0$ and \lam\ masses in MeV/c$^2$, and
the corresponding offsets in $\Xi$ mass, together with the final,
resulting $\Xi^\pm$ mass values. The errors of the corrected
K$^0$ and \lam\ masses include the spread from the simulation smearing,
see text.}
\end{table}

The widths of the \lam\ and K$^0_s$ mass distributions are somewhat larger 
for data than for simulation. 
A study was made in which the reconstructed variables in simulation, 
one by one, were artificially ``smeared'' and a corresponding extra
measurement error added, such that the width of the mass peak in simulation 
agreed with that in the data. The spreads of the shifts obtained by smearing
different variables, amounting to 0.05 MeV/c$^2$ for K$^0$ and 0.04 MeV/c$^2$ 
for \lam, were included in the errors for the corrected offsets quoted in
Table~\ref{table:v0masses}. However, the means 
of the mass values from the smearings agreed with the ``unsmeared'' values.


Offsets of the corrected \lam\ and K$^0_s$ mass values from their known values
can arise from an error in the correction for dE/dx losses, 
an error in the assumed magnetic field, 
or, most likely, a combination of the two. 

The $\Xi$ mass offset can be expressed as a function of the K$^0_s$
and \lam\ mass offsets, as
$$\Delta M_\Xi = \pmatrix{ b_\Xi & d_\Xi }
\pmatrix{ b_K & d_K \cr b_\Lambda & d_\Lambda }^{-1}
\pmatrix{ \Delta M_K \cr \Delta M_\Lambda }$$
where $b_i$ are the $\Xi$, K$^0_s$ and \lam\ mass shift coefficients due to
magnetic field changes, and $d_i$ are those due to dE/dx correction changes. The
values of these coefficients were found using Monte Carlo techniques:
$b_\Xi = 0.0805\pm0.0004$, $b_K = 0.2376\pm0.0010$, $b_\Lambda =
0.0438\pm0.0002$, $d_\Xi = 0.20\pm0.02$, $d_K = 0.367\pm0.017$ and
$d_\Lambda = 0.162\pm0.011$.



Inserting the observed K$^0_s$ and \lam\ mass shifts and the mass offset
coefficients into the above equation, taking all errors into account,
gave the $\Xi$ mass offsets presented in
Table~\ref{table:v0masses}. The last line of that table gives
the final corrected $\Xi^\pm$ mass values per year.
The final average corrected mass value is
($1321.71\pm0.06\pm0.04$) MeV/c$^2$ 
with a $\chi^2$ probability for the combination of 53\%. 
The second error quoted is the unfolded contribution from the uncertainty
in the calibration offsets.


\subsubsection{Other systematic uncertainties}

The effect of using different parameterizations for the shape of the $\Xi$ 
mass peak and for the background was studied, 
as well as that of using various fitting techniques 
(maximum likelihood and minimum $\chi^2$). 
These variations gave a spread in the final $\Xi$ mass value 
of $\pm0.03$ MeV/c$^2$.

Applying the same ``smearing'' technique to the simulated $\Xi$ events as was
described above for the \lam\ and K$^0_s$ analysis gave a spread in the
final $\Xi$ mass value of $\pm0.02$ MeV/c$^2$. Again the average of the 
values from the smearing study agreed with the unsmeared value.

As a cross-check,
the $\Xi$ mass was measured as a function 
of the momentum of the pion from the \xin\ decay.
This is generally the one passing through the most material,
and it is not affected by the $\Lambda$ mass constraint.
Thus it is the one most sensitive to dE/dx corrections. 
No systematic effect depending on the pion momentum was observed.
The $\Xi$ mass was also measured as a
function of the polar angle of the
$\Xi$ momentum and of the observed distance in the $xy$ plane from the beam
axis.
Again no systematic variation was found.

The total systematic error was thus $\pm0.05$ MeV/c$^2$, as shown in Table
\ref{tab:syst}.

\begin{table}
\begin{center}
{
\begin{tabular}{|l|c|}
\hline               
Source                       & MeV/c$^2$ \\ 
\hline  
\hline
\lam\ and K$^0$ mass scale   & $\pm0.04$ \\
Fit parameterization         & $\pm0.03$ \\
Simulation smearing          & $\pm0.02$ \\
\hline  
total                        & $\pm0.05$ \\
\hline
\end{tabular}
}
\end{center}
\caption{\label{tab:syst} Systematic error contributions to the $\Xi$ mass
                          measurement.}
\end{table}


\subsubsection{Results}

The measured average $\Xi$ masses are:
\begin{eqnarray*}
{\mathrm M}_{\Xi^-} & = & (\wffe{1321.70}{0.08}{0.05}) ~{\mathrm{MeV/c}}^2 \\ 
{\mathrm M}_{\Xi^+} & = & (\wffe{1321.73}{0.08}{0.05}) ~{\mathrm{MeV/c}}^2 \\ 
{\mathrm M}_{\Xi^- + \Xi^+} & = & (\wffe{1321.71}{0.06}{0.05}) ~{\mathrm{MeV/c}}^2,
\end{eqnarray*}
where the systematic errors quoted are common to all three values.

The systematic errors cancel in the mass difference\footnote{
It was checked that there was no difference between masses of \lam\ and
$\overline{\Lambda}$, and that the K$^0$ mass did not depend on the charge
of the highest momentum particle in the decay.
}, where
the small statistical errors on the uncorrected values 
can therefore be fully exploited. 
The mass difference measured in the data is 
\begin{center}
$ \Delta_{\mathrm M} = {\mathrm M}_{\Xi^-} - {\mathrm M}_{\Xi^+}
                 = (-0.03 \pm 0.12)~{\mathrm{MeV/c}}^2$
\end{center}
which corresponds to a fractional mass difference of
\begin{center}
$ ({\mathrm M}_{\Xi^-} - {\mathrm M}_{\Xi^+}) / {\mathrm M}_{average} =
  (-2.5 \pm 8.7) \times 10^{-5}. $
\end{center}
This improves the precision on this CPT violation test quantity by a
factor of 3 compared to the current PDG value of $(11 \pm 27)
\times 10^{-5}$~\cite{PDG}.


\subsection{Measurement of \xin\ and \xip\ lifetimes and lifetime difference}

The measurement of the mean lifetimes of the {\xin} and {\xip} and their
lifetime differences uses
the {\xin} and {\xip} candidates with a $\Lambda\pi$ invariant mass within 
$\pm$5 MeV/c$^2$ of the nominal mass, where the signal to background ratio
is about 6:1. 
This is the sample for which data and simulation were compared in detail in
Figure~\ref{fig:mc_data_xi_1}.

\begin{figure}
\begin{center}
\mbox{\epsfig{figure=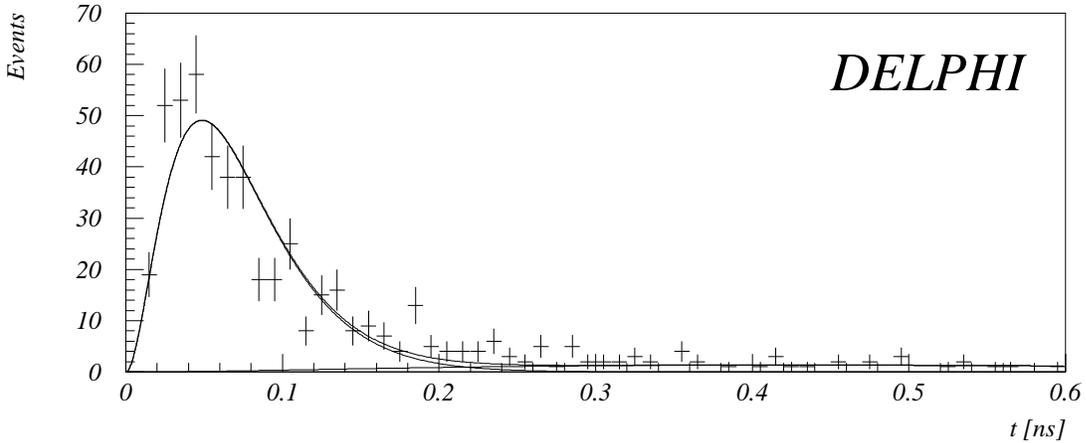,width=\textwidth}} 
\end{center}
\caption{The observed time distribution in the wrong-sign sample for 1992-1995
data. The two lower curves are the $b$-functions described in the text. Their 
sum, used to describe the combinatorial background, is also shown. Only events
with times larger than 0.04~ns were used in the fit.}
\label{fig:tauws}
\end{figure}

\begin{figure}
\begin{center}
\mbox{\epsfig{figure=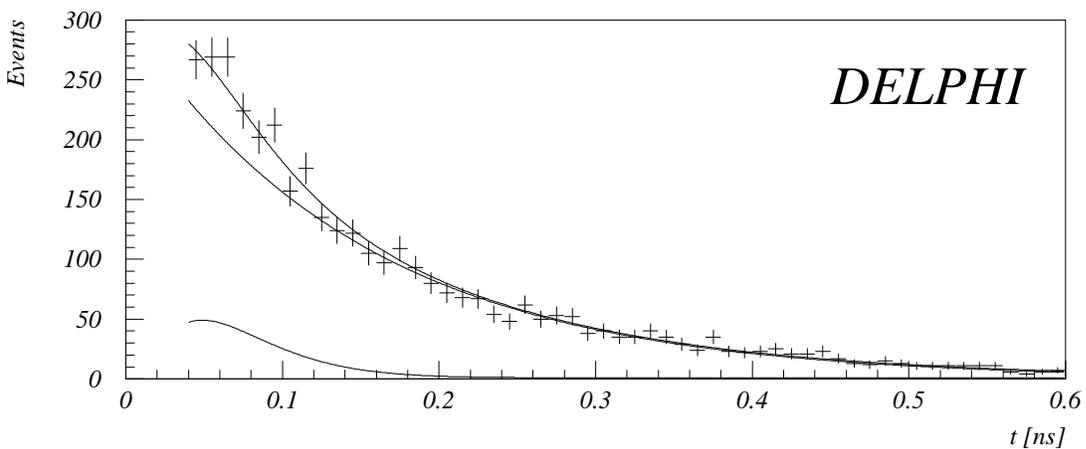,width=\textwidth}} 
\end{center}
\caption{The observed time distribution in the right-sign sample for 1992-1995
data. The lowest curve is the estimate of the contribution from combinatorial
background events, obtained by fitting the wrong-sign combinations.
The middle curve is the estimate of the contribution of \xin\ and \xip\ decays. 
The upper curve represents the fit to the observed time distribution,  
{\it i.e.} the sum of the two lower distributions.}
\label{fig:taurs}
\end{figure}

The mean lifetimes were estimated using a maximum likelihood fit. The time
distribution of the combinatorial background was estimated simultaneously 
in the fit by using the wrong-sign combinations. 
The observed proper time distributions
and the fitted functions for the
wrong-sign and right-sign distributions are shown in Figures~\ref{fig:tauws}
and \ref{fig:taurs} respectively. As the mean lifetimes are much 
shorter for $c$- and $b$-baryons than for a $\Xi$, 
all $\Xi$ may safely be assumed to originate from the interaction point. 


The proper time was calculated as 
\begin{eqnarray}
t & = & {d_{\Xi}M_{\Xi} / P_{\Xi}}
\end{eqnarray}
where $d_{\Xi}$ is the fitted flight distance in the $xy$ plane,
$P_{\Xi}$ is the fitted momentum of the $\Xi$ candidate
in the $xy$ plane, and
$M_{\Xi}$ is the nominal $\Xi$ mass.

For right-sign and wrong-sign candidates in the proper time 
interval 0.04 ns to 2.0 ns,
the following likelihood function was formed:
\begin{eqnarray}
  {\cal L} & = & \prod_{i=1}^{N_{rs}} F(t_i)\ \cdot 
                 \prod_{j=1}^{N_{ws}} B(t_j).
\end{eqnarray}
The first factor, the $F(t)$ product, represents
the right-sign ($\Lambda\pi^-$, $\overline{\Lambda}\pi^+$) 
combinations. The second factor, the $B(t)$ product, 
is an empirical parameterization of the wrong-sign 
($\Lambda\pi^+$, $\overline{\Lambda}\pi^-$) combinations. 
The same function $B(t)$ was also used to describe the background 
in the right-sign sample.
Thus, by maximizing the joint 
likelihood function $\cal L$, the background contribution in the right-sign
sample was naturally constrained to
the shape of the wrong-sign distribution.

The right-sign function $F(t)$ was given by
\begin{eqnarray}
F(t) & = & {1\over\sigma_0+1}\left(\sigma_0 S(t) + B(t)\right)
\end{eqnarray}
where $S(t)$ is a normalized probability density function 
for the observed signal,
{\em i.e.} it is proportional to $\epsilon(t)e^{-{t/\tau_{\Xi}}}$, 
where $\epsilon(t)$ is an empirical efficiency parameterization
of the time-dependent form $e^{(c_1+c_2 t)}$ determined from the simulation.
The relative normalization of the signal $S(t)$ and background $B(t)$ 
in the right-sign sample, $\sigma_0$, was fixed
by the observed number of right-sign ($N_{rs}$) and wrong-sign ($N_{ws}$) 
events in the fitted time interval 0.04 ns to 2.0 ns,
$\sigma_0 = {N_{rs}-N_{ws}\over N_{ws}}$. 


The background function $B(t)$ was given by
\begin{eqnarray}
B(t) & = & {1\over b_1+1} \left\{ b_1{b(t;\sigma_1)\over{\cal N}_1}
                                  +{b(t;\sigma_2)\over{\cal N}_2} \right\}
\end{eqnarray}
where 
\begin{eqnarray}
b(t;\sigma_i) & = & {1\over \Gamma(\beta)\sigma_i} 
          \left({t\over\sigma_i}\right)^{\beta-1} e^{-{t\over\sigma_i}}.
\label{eqn:beta}
\end{eqnarray}
and ${\cal N}_1$ and ${\cal N}_2$ are normalization 
constants for the two $\Gamma$-distributions $b(t;\sigma_i)$.
The value $\beta=3$ provided a good description
of the wrong-sign distribution.
The  parameters $b_1$, $\sigma_1$ and $\sigma_2$ were fitted to the data,
together with $\tau_{\Xi}$. The fit results for each year are
given in Table~\ref{tab:tau_yr}.

\begin{table}
\begin{center}
\begin{tabular}{|c|rrrr|}
\hline               
Year  & \multicolumn{1}{c}{92} & \multicolumn{1}{c}{93} &
        \multicolumn{1}{c}{94} & \multicolumn{1}{c|}{95} \\
\hline
\hline
$\tau_{\Xi^-}$ & $0.131\pm0.012$ & $0.179\pm0.014$ & $0.199\pm0.015$
               & $0.167\pm0.016$ \\
$\tau_{\Xi^+}$ & $0.165\pm0.015$ & $0.146\pm0.013$ & $0.205\pm0.015$
               & $0.169\pm0.020$ \\
\hline
$\Delta\tau=\tau_{\Xi^-} - \tau_{\Xi^+}$ & $-0.034\pm0.020$ & $+0.033\pm0.019$
               & $-0.006\pm0.021$ & $-0.002\pm0.025$ \\
\hline
\end{tabular}
\end{center}
\caption{\label{tab:tau_yr} Fit results and statistical errors for \xin\ and
\xip\ lifetime fits. Values are in nanoseconds.} 
\end{table}

The measured \xin\ and \xip\ lifetimes are: 
\begin{eqnarray*}
 \tau_{\Xi^-} & = & (\wffe{0.165}{0.007}{0.012}) ~{\mathrm{ns}} \\
 \tau_{\Xi^+} & = & (\wffe{0.170}{0.008}{0.012}) ~{\mathrm{ns}} \\
 \tau_{\Xi^- + \Xi^+} & = & (\wffe{0.167}{0.006}{0.012}) ~{\mathrm{ns}} 
\end{eqnarray*}
where the results were achieved by performing the same analysis
on the four separate years. The lifetimes were taken as the
weighted average of the four years.

In order to minimize the effect of statistical fluctuations, the combined
\xin\ and \xip\ sample was used to evaluate the systematic errors.
The following
sources of systematic errors were considered:
\begin{itemize}
\item     the effect on the lifetime fits of the uncertainty in
          the parameters $c_1$ and $c_2$ in the efficiency 
          parameterization was estimated using the simulation
          to be $\pm$0.005 ns;
\item  
          the difference between the input and reconstructed lifetimes
          in the simulation was ($0.002\pm$0.004) ns;
\item  
          changing the fit range between 0.04 ns and 0.08 ns
          for the lower boundary and between 0.6 ns and 2.0 ns for the
          upper boundary changed the fitted lifetime by $\pm$0.004 ns;
\item 
          changing the value of $\beta$ (Equation~\ref{eqn:beta}) in the
          range 1.5 to 4.0 had no significant
          influence on the final results;
\item  
          the $\chi^2$ of the combination of the four
          different years was 14.1 for 3 degrees of freedom corresponding
          to a probability of only 0.3\%; applying the PDG scaling procedure 
          to the combination gives an additional systematic error of
          $\pm0.011$ ns.
\end{itemize}

Note that a $\Xi$ could be reconstructed only if the (anti)proton 
from the (anti)lambda was seen in the TPC. Therefore the fact that 
more antibaryons than baryons interacted in the material before the TPC reduced
the relative number of \xip\ reconstructed by about 10\%,
but had no significant effect on their lifetime distribution.
Thus the systematic errors quoted above are common to all three lifetime values.

The systematic errors cancel in the measurement 
of the lifetime difference, averaged over the years:
\begin{center}
$<\Delta\tau> = <\tau_{\Xi^-} - \tau_{\Xi^+} >
             = (-0.002 \pm 0.011$) ns,
\end{center}
which gives a fractional lifetime difference of
\begin{center}
$ ({\tau}_{\Xi^-} - {\tau}_{\Xi^+}) / {\tau}_{average} = -0.01 \pm 0.07.$
\end{center}
This quantity, which would indicate violation of CPT invariance
if different from zero, has a much smaller error than the PDG value of
$0.02 \pm 0.18$~\cite{PDG}.

The value of $ \Delta\tau $ may also be used together with the world average
for the \xin\ lifetime, \mbox{$\tau_{\Xi^-}^{\mathrm{PDG}} = 
(0.1639 \pm 0.0015)$ ns},
to make a new precise determination of the \xip\ lifetime alone:
\begin{center}
$ \tau_{\Xi^+} = \tau_{\Xi^-}^{\mathrm{PDG}} - \Delta\tau
                = (0.166 \pm 0.011)$ ns.
\end{center}

\subsection{Measurements of \xin\ and \xip\ production}

The parameterization of the signal used in the $\Xi$ mass determination
was not used
to evaluate the efficiencies and production rates,
since the broader Gaussian 
tended to become unreasonably wide 
if left free when fitting substantially smaller data samples. 
Instead, as in the lifetime analysis just described,
a fixed interval, this time $\pm$10 MeV/c$^2$ around the nominal {\xin}
mass, was used as signal region.
The background was estimated from the
wrong-sign invariant mass
distributions. 

\begin{figure}
\begin{center}
\mbox{\epsfig{figure=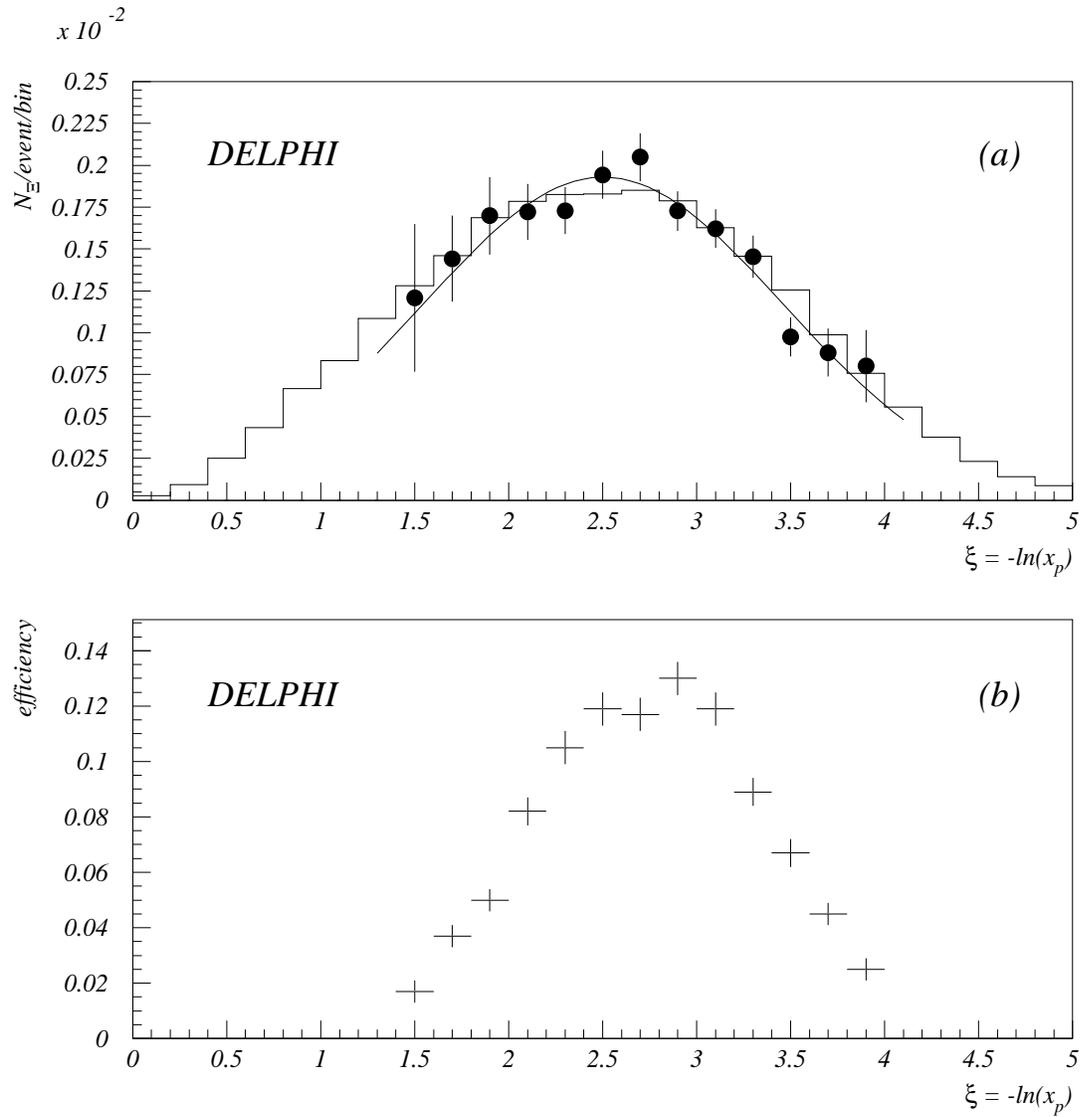,width=\textwidth}}
\end{center}
\caption{{\bf(a)} Efficiency-corrected distribution of $\xi = -\ln x_p$:
the points
with error bars represent the measured $\xi$ distribution, a fit to a Gaussian
function is superimposed, and the \jetset\ $\xi$ spectrum is shown
as the solid histogram. {\bf(b)} The \xin\ reconstruction efficiency 
as a function of $\xi$.}
\label{fig:ksi}
\end{figure}

The efficiency was determined from simulation and depended on the $\Xi$ momentum
(see Table~\ref{table:xieff} and Figure~\ref{fig:ksi}).
The average efficiency was found to be (6.76~$\pm$~0.27~(stat.))\%
for the combined {\xin} and {\xip} reconstruction, including
cuts and the 63.9\%
branching ratio for $\Lambda \rightarrow$ p$\pi^-$. 
The error comes from the finite number of simulated events. 
As mentioned earlier, the reconstruction efficiency was $\sim 10$\%
lower for {\xip} than for \xin, due to differences in the 
cross-sections for hadronic interactions 
of particles and antiparticles in the detector material.


All {\xin} candidates satisfying the standard $\Xi^-$ cuts, and 
with a $\Lambda\pi^-$ invariant mass within $\pm$10~MeV/c$^2$ of the 
nominal {\xin} mass, were considered. The background contribution was estimated
from the wrong-sign combinations and was subtracted.
The measured distribution in $\xi = -\ln x_p$ is shown in Figure~\ref{fig:ksi}
and Table~\ref{table:xieff}.
The $ < \xin + \xip >$ production rate 
in the $\xi$ interval $1.4 < \xi < 4.0$ was found to be
\begin{center}
$ < \xin + \xip >_{q\bar q} = 0.0197 \pm 0.0007~{\mathrm{(stat.)}}$
\end{center}
where the statistical error includes the contributions from data
and simulation.
Extrapolating to the full momentum range
using the \jetset prediction gave
\begin{eqnarray*}
& < \xin + \xip >_{q\bar q} & = \wffe{0.0247}{0.0009}{0.0025}
\end{eqnarray*}
in hadronic \zz\ decays.
This result agrees with the previous DELPHI value of
$0.0250 \pm 0.0009 \pm 0.0021$~\cite{spyros},
obtained using a somewhat different $\Xi$ reconstruction procedure,
and with the OPAL value of $0.0259 \pm 0.0004 \pm 0.0009$~\cite{opal_zpc}.
For comparison, the DELPHI tuned \jetset 7.3~\cite{tuning} gives 0.0251 
and \jetset 7.4 with default parameters gives 0.0273, 
whereas \herwig 5.9~\cite{HERWIG} gives 0.0730.

The systematic error quoted above has the following two sources.
Firstly, according to the simulation, 20\% of the 
{\xin} and {\xip} were produced
outside the range $1.4  < \xi < 4.0$. 
An error of 50\% of this number was taken as a contribution to the total 
systematic error. 
Secondly, adding a cut on lifetime, $\tau_{\Xi} > 0.1$ ns, and requiring 
the \lam\ candidate to be tagged as a `tight' \lam\ ($xy$ flight distance 
above four standard deviations and $\chi^2$ probability 
larger than 1\%) by the $V^0$ reconstruction
program gives a very clean sample.
The production rate calculated with this sample, extrapolated to the
full momentum range,
is $0.0258 \pm 0.0012(\mathrm{stat.})$, which is
the same as that above within errors.
The half-difference of the rates calculated with the two different
sets of cuts was added in quadrature to give the total systematic error.
The effect of varying the width of the signal region was very small.


From a Gaussian fit in the interval $1.4 < \xi < 4.0$, the $\xi$ distribution
was found to have a maximum at
\begin{center}
$ \xi_{data}^* = \wffe{2.50}{0.06}{0.04}$
\end{center}
where the systematic error was evaluated by varying the fit region and the
$\Xi$ mass window.
The \jetset model, 
with parameters tuned as in
\cite{tuning}, gave $\xi_{JETSET}^* =$ 2.522 $\pm$ 0.004(stat.) from a 
similar fit.
These values are slightly lower than the OPAL measurement of $ \xi^* =$ 2.72
$\pm$ 0.13~\cite{opal_zpc}.

The large statistics of the \jetset {\xin} sample 
clearly showed that the generated $\xi$ distribution was not Gaussian.
However, fitting a modified Gaussian form~\cite{mlla}
to the generated $\xi$ spectrum gave $\xi^* = 2.506 \pm 0.004$, 
very close to the result of fitting the unmodified Gaussian. 
Fitting the same modified form to the data, keeping the
skewness and kurtosis parameters fixed to the values found in the simulation,
gave $\xi^* = 2.51 \pm 0.06$(stat). These fits were also to the region
$1.4 < \xi < 4.0$.

\begin{table}
\begin{center}
{
\begin{tabular}{|c|c|c|c|c|c|}
\hline
Momentum          &
       $x_p$      &
 $\xi = -\ln x_p$ &
Efficiency        &
Reconstructed     & 
$N_{\Xi}$/bin/event \\
(GeV/c)           &
                  &
                  &
(\%)              &
      $\Xi^-$     & 
                 \\
\hline
\hline
  ~9.21--11.24 
& 0.202--0.246 
& 1.4--1.6 
& 1.7$\pm$0.4 
&  65$\pm$17  
& 0.00118$\pm$0.00042 \\
  7.54--9.21 
& 0.165--0.202 
& 1.6--1.8 
& 3.7$\pm$0.4 
& 173$\pm$21 
& 0.00145$\pm$0.00026 \\
  6.17--7.54 
& 0.135--0.165 
& 1.8--2.0 
& 5.0$\pm$0.4 
& 290$\pm$28 
& 0.00178$\pm$0.00024 \\
  5.05--6.17 
& 0.111--0.135 
& 2.0--2.2 
& 8.2$\pm$0.5 
& 474$\pm$33 
& 0.00177$\pm$0.00016 \\
  4.14--5.05 
& 0.091--0.111 
& 2.2--2.4 
& 10.5$\pm$0.6 
& 604$\pm$35 
& 0.00180$\pm$0.00014 \\
  3.39--4.14  
& 0.074--0.091 
& 2.4--2.6 
& 11.9$\pm$0.6 
& 758$\pm$37 
& 0.00196$\pm$0.00014 \\
  2.77--3.39 
& 0.061--0.074 
& 2.6--2.8 
& 11.7$\pm$0.6 
& 804$\pm$38 
& 0.00210$\pm$0.00014 \\
  2.27--2.77 
& 0.050--0.061 
& 2.8--3.0 
& 13.0$\pm$0.6 
& 747$\pm$36 
& 0.00176$\pm$0.00011 \\
  1.86--2.27 
& 0.041--0.050 
& 3.0--3.2 
& 11.9$\pm$0.6 
& 633$\pm$31 
& 0.00165$\pm$0.00011 \\
  1.52--1.86 
& 0.033--0.041 
& 3.2--3.4 
& 8.9$\pm$0.5 
& 433$\pm$27 
& 0.00151$\pm$0.00012 \\
  1.25--1.52 
& 0.027--0.033 
& 3.4--3.6 
& 6.7$\pm$0.5 
& 221$\pm$20 
& 0.00101$\pm$0.00012 \\
  1.02--1.25 
& 0.022--0.027 
& 3.6--3.8 
& 4.5$\pm$0.4 
& 133$\pm$16 
& 0.00092$\pm$0.00014 \\
  0.84--1.02
& 0.018--0.022
& 3.8--4.0
& 2.5$\pm$0.4
& 64$\pm$12
& 0.00081$\pm$0.00020 \\
\hline
\end{tabular}
}
\end{center}
\caption{\label{table:xieff} $\Xi^-$ efficiency and $\xi$ distribution.
}
\end{table}


The above procedure for finding \xin\ 
was also applied to $Z\to b\bar b$ decays.
The $b\bar b$ events were selected
with a lifetime tag algorithm~\cite{NIM2,btagd}. 
This technique is based on the measurement of
the impact parameter of each particle relative to the $Z^0$ production point.
Decay products from particles with relatively
long lifetimes, like $B$-hadrons, will have large impact parameters.
Particles produced in the primary
interaction will have impact parameters with a 
spread around zero according to the 
spatial resolution of the detector. 
From all tracks with a positive impact parameter in an event,
the probability for the hypothesis that they all came from a single point 
was calculated.
Events in which this probability was below 1\% 
were selected as $b\bar b$ events.
The joint efficiency to reconstruct a ${\xin}$ decay
and simultaneously tag a $b\bar b$ event with this cut 
was about 3\%, with a $b\bar b$ purity of 77\%. Using these
results the production rate of \xin\ and \xip\ in $b\bar b$ events
was calculated. 
Taking the weighted average of the results from the four years gives the 
final rate: 
\begin{center}
$ < {\xin} + {\xip} >_{b\bar b} = \wffe{0.0183}{0.0016}{0.0035}$
\end{center}
where the systematic error comes from the momentum extrapolation
and the sample variation of the 
four years' data.
Different cuts on the probability 
as well as looser ${\xin}$ selections were also tested. The value of 
$< {\xin} + {\xip} >_{b\bar b}$ changed only within $\pm 0.001$.
The DELPHI tuned \jetset 7.3~\cite{tuning} gives $< \xin + \xip >_{b\bar b}$
=0.0238 and \jetset 7.4 with default parameters gives 0.0208, whereas
\herwig 5.9 gives $< \xin + \xip >_{b\bar b}$ = 0.0523.

\section{Summary}

About 2500 \xin\ and 2300 \xip\ decays have been reconstructed from data
collected by the DELPHI detector in the years 1992 to 1995. 

From this large sample, direct measurements have been made of the
\xin\ and \xip\ masses and their average and difference:
\begin{eqnarray*}
 {\mathrm M}_{\Xi^-} & = &(\wffe{1321.70}{0.08}{0.05}) ~{\mathrm{MeV/c}}^2 \\ 
 {\mathrm M}_{\Xi^+} & = &(\wffe{1321.73}{0.08}{0.05}) ~{\mathrm{MeV/c}}^2 \\ 
 {\mathrm M}_{\Xi^- + \Xi^+} & = &(\wffe{1321.71}{0.06}{0.05}) ~{\mathrm{MeV/c}}^2 \\
 {\mathrm M}_{\Xi^-} - {\mathrm M}_{\Xi^+} & =& (-0.03 \pm 0.12) ~{\mathrm{MeV/c}}^2 \\
({\mathrm M}_{\Xi^-} - {\mathrm M}_{\Xi^+}) / {\mathrm M}_{average} & =&
 (-2.5 \pm 8.7) \times 10^{-5}.
\end{eqnarray*}
The masses given by the PDG~\cite{PDG} are
M$_{\Xi^-}$ = (1321.34 $\pm$ 0.14) MeV/c$^2$,
M$_{\Xi^+}$ = (1321.20 $\pm$ 0.33) MeV/c$^2$ and
(M$_{\Xi^-}$ - M$_{\Xi^+}$) / M$_{average}$ = (11 $\pm$ 27) $\times 10^{-5}$.
Up to now only small samples of \xip\ are referenced by the PDG. 

The \xin\ lifetime measurement obtained is consistent with the PDG value
of (0.1639 $\pm$ 0.0015) ns but has a much larger error.
The lifetime difference obtained:
\begin{center}
$ \Delta\tau = \tau_{\Xi^-} - \tau_{\Xi^+} 
             = (-0.002 \pm 0.011)$ ns
\end{center}
implies
\begin{center}
 $({\tau}_{\Xi^-} - {\tau}_{\Xi^+}) / {\tau}_{average} = -0.01 \pm 0.07$
\end{center}
and, using the PDG value for $\tau_{\Xi^-}$,
\begin{center}
 $\tau_{\Xi^+} = ({0.166} \pm {0.011}) ~{\mathrm ns}$,
\end{center}
which is more precise than our direct measurement, 
$ \tau_{\Xi^+}  =  (0.170\pm0.008\pm0.012)~{\mathrm{ns}}$.
The present PDG values~\cite{PDG} 
for the fractional lifetime difference and
for the \xip\ lifetime are
$({\tau}_{\Xi^-}-{\tau}_{\Xi^+}) / {\tau}_{average} = 0.02 \pm 0.18$
and $\tau_{\Xi^+} = (0.16 \pm 0.03)$ ns respectively.

Thus this analysis significantly improves the precision on the
fractional mass and lifetime
differences of \xin\ and \xip, which test CPT invariance,
compared to the present PDG values.


The inclusive production rates for \xin\ plus \xip\ 
in hadronic $Z$ decays and in $Z\to b\bar b$ decays 
were found to be:
\begin{eqnarray*}
 < {\xin} + {\xip} >_{q\bar q} & = &\wffe{0.0247}{0.0009}{0.0025} \\
 < {\xin} + {\xip} >_{b\bar b} & = &\wffe{0.0183}{0.0016}{0.0035}
\end{eqnarray*}
The \jetset predictions agree with these measurements,
whereas the \herwig predictions do not.
The maximum of the $\xi = -\ln x_p$ distribution was found to be at
\begin{center}
$ \xi^* = \wffe{2.50}{0.06}{0.04}$
\end{center}
in hadronic $Z$ decays.

\subsection*{Acknowledgements}
\vskip 3 mm
We are greatly indebted to our technical 
collaborators, to the members of the CERN-SL Division for the excellent 
performance of the LEP collider, and to the funding agencies for their
support in building and operating the DELPHI detector.\\
We acknowledge in particular the support of \\
Austrian Federal Ministry of Education, Science and Culture,
GZ 616.364/2-III/2a/98, \\
FNRS--FWO, Flanders Institute to encourage scientific and technological 
research in the industry (IWT), Belgium,  \\
FINEP, CNPq, CAPES, FUJB and FAPERJ, Brazil, \\
Czech Ministry of Industry and Trade, GA CR 202/99/1362,\\
Commission of the European Communities (DG XII), \\
Direction des Sciences de la Mati$\grave{\mbox{\rm e}}$re, CEA, France, \\
Bundesministerium f$\ddot{\mbox{\rm u}}$r Bildung, Wissenschaft, Forschung 
und Technologie, Germany,\\
General Secretariat for Research and Technology, Greece, \\
National Science Foundation (NWO) and Foundation for Research on Matter (FOM),
The Netherlands, \\
Norwegian Research Council,  \\
State Committee for Scientific Research, Poland, SPUB-M/CERN/PO3/DZ296/2000,
SPUB-M/CERN/PO3/DZ297/2000, 2P03B 104 19 and 2P03B 69 23(2002-2004)\\
FCT - Funda\c{c}\~ao para a Ci\^encia e Tecnologia, Portugal, \\
Vedecka grantova agentura MS SR, Slovakia, Nr. 95/5195/134, \\
Ministry of Science and Technology of the Republic of Slovenia, \\
CICYT, Spain, AEN99-0950 and AEN99-0761,  \\
The Swedish Research Council,      \\
Particle Physics and Astronomy Research Council, UK, \\
Department of Energy, USA, DE-FG02-01ER41155, \\
EEC RTN contract HPRN-CT-00292-2002. \\



\newpage

%

\newpage

\newpage

\newpage 

%
%

\newpage

\end{document}